\documentstyle[aps,amsmath,manuscript]{revtex}

\begin{document}

\draft

\title{\bf LOCALIZATION FROM CONDUCTANCE IN FEW-CHANNEL DISORDERED WIRES}
\author{J.Heinrichs}
\address{Institut de physique, B5, Universit\'{e} de Li\`{e}ge, Sart
Tilman, B-4000 Li\`{e}ge, Belgium}
\maketitle
\date{\today}
\maketitle

\begin{abstract}
\thispagestyle{empty}
\noindent
We study localization in two- and three channel quasi-1D systems using
multichain tight-binding Anderson models with
nearest-neighbour interchain hopping.  In the three chain case we discuss
both the case of free- and that of periodic boundary
conditions between the chains.  The finite disordered wires are connected
to ideal leads and the localization length is
defined from the Landauer conductance in terms of the transmission
coefficients matrix.  The transmission- and reflection
amplitudes in properly defined quantum channels are obtained from
$S$-matrices constructed from transfer matrices in Bloch
wave bases for the various quasi-1D systems.  Our exact analytic
expressions for localization lengths for weak disorder
reduce to the Thouless expression for 1D systems in the limit of vanishing
interchain hopping.  For weak interchain hopping
the localization length decreases with respect to the 1D value in all
three cases.  In the three-channel cases it increases
with interchain hopping over restricted domains of large hopping.
\pacs{72.10.-d, 73.70.+m, 73.23.-b - e-mail: J.Heinrichs@ulg.ac.be}
\end{abstract}
\newpage

\pagestyle{plain}
\setcounter{page}{1}

\section{INTRODUCTION}

A wire is a topologically one-dimensional\ system whose width of the order
of the square root of the cross-sectional area
$A$  is much smaller than its length $L$ i.e. $\sqrt A << L$.  In a thin
wire the motion of electrons in the transverse
direction is quantized.  The corresponding transverse eigenstates below
the Fermi level define a finite number,
$N\propto\sqrt{2A\over\lambda_F}$ (with $\lambda_F$, the Fermi wavelength), of
quantum channels for transmission of
electrons across the wire.  The starting point of the present work is the
Landauer two-probe conductance formula\cite{1,2}

\begin{equation}\label{1}
g={2e^2\over h}Tr\;(\hat t\hat t^+)\quad ,
\end{equation}

\noindent
which describes current transport in a disordered wire.  Here $\hat t$ is
the so-called transmission matrix of the
$N$-channel system:

\begin{equation}\label{2}
\hat t=
\begin{pmatrix}
t_{11} & t_{12} & \ldots & t_{1N}\\
t_{21} & \ldots & \ldots & \ldots\\
\ldots & \ldots & \ldots\\
t_{N1} & t_{N2} & \ldots & t_{NN}
\end{pmatrix}\quad .
\end{equation}

\noindent
An incoming wave from channel $j$ of an ideal lead at one end of the
disordered wire (of length $L$) has a coefficient
$|t_{ij}|^2$ for transmission into channel $i$ of the lead at the other
end. \par

The correctness of the description of the conductance in terms of
transmission channels has received striking experimental
confirmation\cite{3} in the special case of perfectly transmitting
channels, where (\ref{1}) reduces to

\begin{equation}\label{3}
g={2e^2\over h}N\quad .
\end{equation}

\noindent
These studies relate to quantum point contacts in the form of narrow
conducting two-dimensional strips whose width, hence the
number of discrete transmitting channels, may be varied by varying
externally applied gate voltages.  Conductance steps
corresponding to increasing values of $N$ in (\ref{3}) are clearly
observed\cite{3} .  A further interesting feature of these
experiments is that they offer the possibility of realizing physically not
only purely one-dimensional systems ($N=1$) but
also few-channel systems ($N=2,3,\ldots$) such as those studied below.  We
recall that recent discussions of many-channel
mesoscopic systems have been essentially restricted to the case
$N>>1$\cite{4} i.e. real (metallic) wires whose widths are
much larger than the Fermi wavelength. \par

The behaviour of the conductance in a disordered wire depends strongly on
its length relative to the localization length
$L_c$.  In an infinitely long disordered wire of a given  cross-section
all eigenstates are expected to be localized just like
in the truly one-dimensional case.  Thouless has indeed shown\cite{5,6}
that for $L>L_c$, where

\begin{equation}\label{4}
L_c \sim N\ell\quad ,
\end{equation}

\noindent
$\ell$ being the mean free path, the conductance should fall off
exponentially as $e^{-L/L_c}$, which is a clear manifestation
of  localization.  On the other hand, one the scales of lengths in the
domain

\begin{equation}\label{5}
\ell<L<L_c=N\ell\quad ,
\end{equation}

\noindent
the eigenstates appear as being delocalized.  In fact, it was shown
later\cite{7} that for lengths in the range (5) there
remains a  number $N_{\text{eff}}\sim{N\ell\over L}$ of independent
ballistic channels leading to metallic Ohm's law behaviour
for the conductance (\ref{1}).  Such a diffusive quasi-metallic domain
does not exist for truly one-dimensional systems where
$L_c\equiv L_{1c}\sim\ell$.  Dorokhov\cite{8} has developed a detailed
scaling analysis\cite{9} of localization in a
multichannel wire, in which he calculates $L_c$ in terms of a
phenomenological mean free path entering as input via Ohm's law
at short scales.  His final result, which is valid for weak disorder, is

\begin{equation}\label{6}
L_c=(N+1)\ell\quad ,
\end{equation}

\noindent
which coincides with (\ref{4}) for $N>>1$ and suggests that $L_c$ is not
directly proportional to the number of channels for
few channel systems.   In particular, for one-dimensional systems  it
yields

\begin{equation}\label{7}
L_{1c}=2\ell\quad .
\end{equation}

\noindent
We recall that (\ref{7}) coincides with the result obtained long ago by
Thouless\cite{10} from kinetic transport theory.  The
same result  follows also from the quantum series composition law for
conductances $g=|t|^2$ for 1D conductors, assuming Ohm's
law to be valid at short length scales\cite{11}.  On the other hand, we
note that for two-channel systems (\ref{6}) yields
$L_c\equiv L_{2c}=3\ell$ which is comparable to the result
$L_{2c}=2\ell(1-{1\over \pi})^{-1}$ obtained by Dorokhov\cite{12}
in a different scaling treatment.
\par

Shortly before the development of the scaling theory of localization for
1D conductors\cite{9}, Thouless derived his
well-known  analytic expression for the quantum localization length in a
tight-binding linear chain for weak disorder\cite{13}.  The
purpose of the present work is to derive similar exact microscopic
expressions for the localization length in
quasi-one-dimensional few channel systems, specifically for two- and three
chain systems for weak disorder.  This study is of
interest in several respects.  For example, via the dependence of the
localization length on the energy across the energy
bands of the pure systems, it provides a first principles quantum proof of
the fact that all states in the
quasi-one-dimensional systems are localized.  It is also relevant for
experimental situations e.g. for discussing the
quasi-metallic domain in few channel systems which are
encountered in semiconducting microstructures\cite{14} or
may be fabricated artificially, as in the quantum point contact system
discussed above. \par

In II.a we define the tight-binding two- and three chain systems for
modelling few channel wires.  In II.b we describe our
methodology for studying the localization length.  It consists in
constructing successively transfer- and
transmission-reflection matrices for the tight-binding systems by
generalizing well-known methods for 1D systems.  It further
relies on general results concerning the existence and the properties of a
Lyapunov exponent (inverse localization length)
describing the asymptotic exponential decay of the conductance (\ref{1})
in mutichannel systems\cite{15}.  In Sect. III we
present the details of our calculations leading to the final analytic
expressions for the transfer and scattering matrices for
the two- and three channel wires for weak disorder.  For the case $N=3$,
we obtain different results for open- and for
periodic boundary conditions which correspond to packing the chains on a
plane and on a cylindrical surface, respectively.   The
final analytic expressions for the averaged transmission- and reflection
coefficients and for localization lengths are
discussed in Sect. IV.  For clarity's sake some details of these
calculations are relegated to an appendix.

\section{FEW-CHANNEL WIRES AND LOCALIZATION}
\subsection{Anderson models in channel bases}

We describe two ($N=2$)- and three ($N=3$)-channel wires by Anderson
models for two- and three coupled chain systems,
respectively.  The two chain Anderson model  consists of parallel linear
chains of $N_L$ disordered sites each (of spacing
$a=1$ and length
$L=N_L a$) connected at both ends to semi-infinite ideal (non-disordered)
leads.  It is defined by the tight-binding
Schr\"{o}dinger equation which we write in the matrix form

\begin{equation}\label{8}
\begin{pmatrix}
\varphi^1_{n+1}+\varphi^1_{n-1}\\
\varphi^2_{n+1} +\varphi^2_{n-1}
\end{pmatrix}=
\begin{pmatrix}
E-\varepsilon_{1n} & -h\\
-h & E-\varepsilon_{2n}
\end{pmatrix}
\begin{pmatrix}
\varphi^1_n\\
\varphi^2_n
\end{pmatrix}\quad ,
\end{equation}

\noindent
where the $\varphi^i_m$ denote the wave-function amplitudes at sites $m$
on the chain $i$, $h$ is a constant matrix
element for an electron to  hop transversally between a site $n$ on chain
1 and its nearest-neighbour site $n$ on chain 2.
The site energies $\varepsilon_{im}$ are random variables associated with
the sites $1\leq m\leq N_L$ of the disordered chain
$i$, and
$\varepsilon_{im}=0$ on the semi-infinite ideal chains defined by the
sites $m>N_L$ and $m<1$, respectively.  The above
energies, including $E$, are measured in units of the constant hopping
rate along the individual chains. \par

The coupled three chain ($N=3$) Anderson model is defined in a similar way
by a set of tight-binding Schr\"{o}dinger
equations, whose  actual form depends, however, on interchain boundary
conditions.  For free boundary conditions, which
correspond to arranging the parallel equidistant chains on a plane, the
tight-binding equations are

\begin{equation}\label{9}
\begin{pmatrix}
\varphi^1_{n+1}+\varphi^1_{n-1}\\
\varphi^2_{n+1} +\varphi^2_{n-1}\\
\varphi^3_{n+1} +\varphi^3_{n-1}
\end{pmatrix}=
\begin{pmatrix}
E-\varepsilon_{1n} & -h & 0\\
-h & E-\varepsilon_{2n} & -h\\
0 & -h & E-\varepsilon_{3n}
\end{pmatrix}
\begin{pmatrix}
\varphi^1_n\\
\varphi^2_n\\
\varphi^3_n
\end{pmatrix}\quad ,
\end{equation}

\noindent
with the sites in the disordered sections of length $L=N_L a$ and in the
semi-infinite ideal chain sections
labelled in the same way as  in the two chain case.  On the other hand, in
the case of periodic boundary conditions which
correspond to equidistant linear chains on a cylindrical surface the
Schr\"{o}dinger equation is

\begin{equation}\label{10}
\begin{pmatrix}
\varphi^1_{n+1}+\varphi^1_{n-1}\\
\varphi^2_{n+1} +\varphi^2_{n-1}\\
\varphi^3_{n+1} +\varphi^3_{n-1}
\end{pmatrix}=
\begin{pmatrix}
E-\varepsilon_{1n} & -h & -h\\
-h & E-\varepsilon_{2n} & -h\\
-h & -h & E-\varepsilon_{3n}
\end{pmatrix}
\begin{pmatrix}
\varphi^1_n\\
\varphi^2_n\\
\varphi^3_n
\end{pmatrix}\quad .
\end{equation}

As discussed in Sect. I, a quasi-one-dimensional wire is described by a
collection of independent channels for wave
transmission.  Microscopic models for two- and three channel wires are
obtained from the systems of tight-binding equations
(\ref{8}-\ref{10}) by diagonalizing the interchain coupling terms in the
equations describing the ideal leads.  This indeed
leads to independent quantum channels for the leads defined by new
amplitudes bases

$$\begin{pmatrix}
\vdots\\\psi_n^i\\\vdots\end{pmatrix}
=\widehat U^{-1}
\begin{pmatrix}
\vdots\\\varphi_n^i\\\vdots\end{pmatrix}
$$

\noindent
in which the non-random parts of the
matrices on the r.h.s. of (\ref{8}-\ref{10}) are diagonal.  For the
multichain system above we obtain, respectively.

\begin{eqnarray*}
\begin{pmatrix}
\psi^1_n\\ \psi^2_n
\end{pmatrix}=\widehat U_0
\begin{pmatrix}
\varphi^1_n\\ \varphi^2_n
\end{pmatrix}\;  &,& \;
\widehat U_0={1\over \sqrt 2}
\begin{pmatrix}
1 & 1\\ 1 & -1
\end{pmatrix}\quad ,\nonumber\\
\widehat U_0
\begin{pmatrix}
E & -h\\-h & E
\end{pmatrix}
\widehat U_0 &=&
\begin{pmatrix}
E-h & 0\\ 0 & E+h
\end{pmatrix}\quad ,\hspace{5cm}\text{(8.a)}
\end{eqnarray*}

\begin{eqnarray*}
\begin{pmatrix}
\psi^1_n\\ \psi^2_n \\ \psi^3_n
\end{pmatrix}=\widehat U'
\begin{pmatrix}
\varphi^1_n\\ \varphi^2_n \\ \varphi^3_n
\end{pmatrix}\;  &,& \;
\widehat U'={1\over 2}
\begin{pmatrix}
1 & \sqrt 2 & 1\\ \sqrt 2 & 0 & -\sqrt 2\\ 1 & -\sqrt 2 & 1
\end{pmatrix}\quad ,\nonumber\\
\widehat U'
\begin{pmatrix}
E & -h & 0\\-h & E & -h \\0 & -h & E
\end{pmatrix}
\widehat U' &=&
\begin{pmatrix}
E-\sqrt 2 h & 0 & 0\\ 0 & E & 0\\ 0 & 0 & E+\sqrt 2 h
\end{pmatrix}\quad ,\hspace{4cm}\text{(9.a)}
\end{eqnarray*}

\begin{eqnarray*}
\begin{pmatrix}
\psi^1_n\\ \psi^2_n \\ \psi^3_n
\end{pmatrix}=(\widehat U\prime\prime)^{-1}
\begin{pmatrix}
\varphi^1_n\\ \varphi^2_n \\ \varphi^3_n
\end{pmatrix}\;  &,& \;
\widehat U"=
\begin{pmatrix}
1 & 1 & 0\\ 1 & 0 & 1\\ 1 & -1 & -1
\end{pmatrix}\quad ,\nonumber\\
(\widehat U")^{-1}
\begin{pmatrix}
E & -h & -h\\-h & E & -h \\-h & -h & E
\end{pmatrix}
\widehat U" &=&
\begin{pmatrix}
E-2 h & 0 & 0\\ 0 & E+h & 0\\ 0 & 0 & E+h
\end{pmatrix}\quad .\hspace{4cm}\text{(10.a)}
\end{eqnarray*}

\noindent
Note that $\widehat U_0$ and $\widehat U'$ are unitary, while $\widehat
U"$ is not. \par

Finally, in the channel bases defined by (8a-10a) the
tight-binding equations (\ref{8}-\ref{10}) read,
respectively,

\begin{equation*}
\begin{pmatrix}
\psi^1_{n+1}+\psi^1_{n-1}\\ \psi^2_{n+1}+\psi^2_{n-1}
\end{pmatrix}=
\begin{pmatrix}
E-h-{1\over 2}(\varepsilon_{1n}+\varepsilon_{2n}) & {1\over
2}(\varepsilon_{2n}-\varepsilon_{1n})\\
{1\over 2}(\varepsilon_{2n}-\varepsilon_{1n}) & E+h-{1\over
2}(\varepsilon_{1n}+\varepsilon_{2n})
\end{pmatrix}
\begin{pmatrix}
\psi^1_n\\ \psi^2_n
\end{pmatrix}\quad ,\hspace{2cm}\text{(8.b)}
\end{equation*}

\begin{eqnarray*}
& &
\begin{pmatrix}
\psi^1_{n+1}+\psi^1_{n-1}\\ \psi^2_{n+1}+\psi^2_{n-1} \\
\psi^3_{n+1}+\psi^3_{n-1}
\end{pmatrix}=\nonumber\\
& \qquad &
\begin{pmatrix}
E-\sqrt 2 h-{1\over
4}(\varepsilon_{1n}+2\varepsilon_{2n}+\varepsilon_{3n}) &
{\sqrt 2\over 4}(\varepsilon_{3n}-\varepsilon_{1n})     &
-{1\over 4}(\varepsilon_{1n}-2\varepsilon_{2n}+\varepsilon_{3n})\\
{\sqrt 2\over 4}(\varepsilon_{3n}-\varepsilon_{1n}) &
E-{1\over 2}(\varepsilon_{1n}+\varepsilon_{3n})     &
{\sqrt 2\over 4}(\varepsilon_{3n}-\varepsilon_{1n})\\
-{1\over 4}(\varepsilon_{1n}-2\varepsilon_{2n}+\varepsilon_{3n}) &
{\sqrt 2\over 4}(\varepsilon_{3n}-\varepsilon_{1n})     &
E+\sqrt 2 h-{1\over 4}(\varepsilon_{1n}+2\varepsilon_{2n}+\varepsilon_{3n})
\end{pmatrix}
\begin{pmatrix}
\psi^1_n\\ \psi^2_n \\ \psi^3_n
\end{pmatrix}\quad ,\\
& &\hspace{15cm}\text{(9.b)}
\end{eqnarray*}

\begin{eqnarray*}
& &
\begin{pmatrix}
\psi^1_{n+1}+\psi^1_{n-1}\\ \psi^2_{n+1}+\psi^2_{n-1} \\
\psi^3_{n+1}+\psi^3_{n-1}
\end{pmatrix}=\\
& &
\begin{pmatrix}
E-2 h-{1\over 3}(\varepsilon_{1n}+\varepsilon_{2n}+\varepsilon_{3n}) &
-{1\over 3}(\varepsilon_{1n}-\varepsilon_{3n})     &
-{1\over 3}(\varepsilon_{2n}-\varepsilon_{3n})\\
-{1\over 3}(2\varepsilon_{1n}-2\varepsilon_{2n}-\varepsilon_{3n}) &
E+h-{1\over 3}(\varepsilon_{1n}+\varepsilon_{3n})     &
{1\over 3}(\varepsilon_{2n}-\varepsilon_{3n})\\
{1\over 3}(\varepsilon_{1n}-2\varepsilon_{2n}+\varepsilon_{3n}) &
{1\over 3}(\varepsilon_{1n}-\varepsilon_{3n})     &
E+h-{1\over 3}(2\varepsilon_{2n}+\varepsilon_{3n})
\end{pmatrix}
\begin{pmatrix}
\psi^1_n\\ \psi^2_n \\ \psi^3_n
\end{pmatrix}\quad ,\\
& &\hspace{15cm}\text{(10.b)}
\end{eqnarray*}

\noindent
which constitute our starting point for deriving transmission- and
reflection matrices of the disordered wires in Sect. III.
It is seen that the similarity transformation of the disorder matrices  by
the $\widehat U$-matrices leads to interchannel
coupling in the disordered sections, $1\leq N\leq N_L$.

\subsection{Localization from conductance}

In Sect. IV we will calculate the localization length in the above
multi-channel wire models from the rate of exponential
decrease (Lyapunov exponent) of the conductance (\ref{1}) for the large
$L$\cite{15}.  The transmission matrix in (\ref{2})
will be found by constructing a transfer matrix which transforms
propagating waves in the multichannel leads (defined by
(8.a-10.a) on the left side of the disordered wire into
corresponding propagating waves on the right side.  The
transfer matrix for the wire of length $L$ is expressed, as usual, as a
product of $N_L$ transfer matrices for small sections enclosing only the
$n$th site of each one of the channels.
The calculation of the localization length rests on
theorems of Oseledec\cite{16} and of Tutubalin and Vister\cite{17} on the
properties of products of a large number of random
matrices.  Indeed, employing these properties Johnston anf Kunz\cite{15}
have shown that the Lyapunov exponent $\gamma$ exists
for the conductance (\ref{1}) and is a self-averaging quantity referred to
as the inverse localization length.  It is defined
by the relation

\begin{equation}\label{11}
\gamma\equiv {1\over L_c}=-\lim_{N_L\rightarrow\infty}{1\over 2N_L}\langle
\ln g\rangle\quad ,
\end{equation}

\noindent
where $\langle\ldots\rangle$ denotes averaging over the disorder (i.e. the
random site energies in the Anderson
model)\cite{18}.  It follows that the  asymptotic distribution of the
conductance is log-normal.

\section{DETAILED ANALYSIS}

As indicated above the transmission matrices of the form (\ref{2}) for the
quasi 1D-disordered systems above will be
obtained from transfer matrices for wavefunction amplitudes defined from
equations (8.b-10.b), respectively.  The
construction of these transfer matrices proceeds in two~steps.  First, we
will define transfer matrices for thin slices
enclosing only one site $n$ of each chain in a disordered wire, in a Bloch
plane wave basis.  Next the transfer matrix of a
whole wire of length $L=N_L a$ will be obtained as a product of the
transfer matrices for the $N_L$ individual slices
composing the wire.  We will express it analytically to lowest order in
the effect of a weak disorder.  Finally we obtain
the form of the tight-binding equations describing the transfer of Bloch
wave amplitudes across a whole disordered wire of length $L$, which we then
cast in
the form of scattering equations in order to identify microscopic
scattering matrices of the form (\ref{2}), in terms of
elements of the transfer matrices.  We recall that the transfer matrix
method is well-known in the study of one-dimensional
disordered systems\cite{19,20,21}.  Here we generalize and adapt it in the
case of few channel quasi-1D systems.

\subsection{Transfer matrices}

Transfer matrices $\tilde X_{0n},\;\tilde X'_n$ and $\tilde X_n"$ for thin
slices including a single site $n$ per chain of the
quasi-1D systems described by (8.b-10.b) are defined by
rewrtiting these equations respectively in the forms

\begin{equation}\label{12}
\begin{pmatrix}
\psi^1_{n+1}\\ \psi^1_n\\ \psi^2_{n+1}\\ \psi^2_n
\end{pmatrix}=\tilde X_{0n}
\begin{pmatrix}
\psi^1_n \\ \psi^1_{n-1}\\ \psi^2_n \\ \psi^2_{n-1}
\end{pmatrix}
\quad ,
\end{equation}

\begin{equation}\label{13}
\begin{pmatrix}
\psi^1_{n+1} \\ \psi^1_n \\ \psi^2_{n+1} \\ \psi^2_n \\ \psi^3_{n+1} \\
\psi^3_n
\end{pmatrix}=\tilde Y_n
\begin{pmatrix}
\psi^1_n \\\psi^1_{n-1} \\ \psi^2_n \\ \psi^2_{n-1}  \\ \psi^3_n \\
\psi^3_{n-1}
\end{pmatrix} ,\tilde Y_n\equiv\tilde X'_n,\tilde X_n"
\quad ,
\end{equation}

\noindent
where

\begin{equation}\label{14}
\tilde X_{0n} =
\begin{pmatrix}
E-h-\mu_n & -1 & \nu_n & 0\\
1 & 0 & 0 & 0\\
\nu_n & 0 & E+h-\mu_n & -1\\
0 & 0 & 1 & 0
\end{pmatrix}
\quad ,
\end{equation}

\begin{equation*}
\mu_n =
{1\over 2} (\varepsilon_{1n}+\varepsilon_{2n}), \nu_n={1\over 2}
(\varepsilon_{2n}-\varepsilon_{1n})\quad ,\hspace{8cm}\text{(14.a)}
\end{equation*}

\begin{equation}\label{15}
\tilde X'_n=
\begin{pmatrix}
E-\sqrt 2 h-\mu'_n & -1 &	 \nu'_n & 0 & \tau'_n & 0\\
1 & 0 & 0 & 0 & 0 & 0\\
\nu'_n & 0 & E-\eta'_n & -1 & \nu'_n & 0\\
0 & 0 & 1 & 0 & 0 & 0\\
\tau'_n & 0 & \nu'_n & 0 &  E+\sqrt 2 h-\mu'_n & -1\\
0 & 0 & 0 & 0 & 1 & 0
\end{pmatrix}\quad ,
\end{equation}

\begin{eqnarray*}
\mu'_n={1\over 4}(\varepsilon_{1n}+2\varepsilon_{2n}+\varepsilon_{3n})
&\;,\;&
\nu'_n={\sqrt 2\over 4}(\varepsilon_{3n}-\varepsilon_{1n})\quad,\nonumber\\
\tau'_n=-{1\over 4}(\varepsilon_{1n}-2\varepsilon_{2n}+\varepsilon_{3n})
&\;,\;& \eta'_n={1\over 2}
(\varepsilon_{3n}+\varepsilon_{1n})\quad ,\hspace{6cm}\text{(15.a)}
\end{eqnarray*}

\begin{equation}\label{16}
\tilde X_n"=
\begin{pmatrix}
E- 2 h-\mu_n" & -1 &	 \nu_n" & 0 & \tau_n" & 0\\
1 & 0 & 0 & 0 & 0 & 0\\
\alpha_n" & 0 & E+h-\eta_n" & -1 & -\tau_n" & 0\\
0 & 0 & 1 & 0 & 0 & 0\\
\beta_n" & 0 & -\nu_n" & 0 &  E+h-\theta_n" & -1\\
0 & 0 & 0 & 0 & 1 & 0
\end{pmatrix}\quad ,
\end{equation}

\begin{eqnarray*}
\mu_n"
&=&
{1\over 3} (\varepsilon_{1n}+\varepsilon_{2n}+\varepsilon_{3n})\;,\;
\eta_n"={1\over 3}(2\varepsilon_{1n}+\varepsilon_{3n})\;,\;
\theta_n"={1\over 3}(2\varepsilon_{2n}+\varepsilon_{3n})\quad ,\nonumber\\
\nu_n"
&=&
-{1\over 3}(\varepsilon_{1n}-\varepsilon_{3n})\;,\;
\tau_n"=-{1\over 3}(\varepsilon_{2n}-\varepsilon_{3n})\quad ,\nonumber\\
\alpha_n"
&=&
-{1\over 3}(2\varepsilon_{1n}-\varepsilon_{2n}-\varepsilon_{3n})\;,\;
\beta_n"={1\over
3}(\varepsilon_{1n}-2\varepsilon_{2n}+\varepsilon_{3n})\quad
.\qquad\qquad\qquad\qquad\text{(16.a)}
\end{eqnarray*}

\noindent
The study of disordered wires in terms of reflection and transmission
properties of plane waves requires determination of plane wave bases in
which the transfer matrices for slices $n$ in the leads are diagonal.
Such bases are provided by the Bloch wave solutions for the leads which
are
defined by

\begin{equation}\label{17}
\tilde X_{00}
\begin{pmatrix}
\psi^1_{n,\pm}\\
\psi^1_{n-1,\pm}\\
\psi^2_{n,\pm}\\
\psi^2_{n-1,\pm}
\end{pmatrix}=
\begin{pmatrix}
e^{\pm ik_1}\psi^1_{n,\pm}\\
e^{\pm ik_1}\psi^1_{n-1,\pm}\\
e^{\pm ik_2}\psi^2_{n,\pm}\\
e^{\pm ik_2}\psi^2_{n-1,\pm}
\end{pmatrix}\quad ,
\end{equation}

\begin{equation}\label{18}
\tilde Y_0
\begin{pmatrix}
\psi^1_{n,\pm}\\
\psi^1_{n-1,\pm}\\
\psi^2_{n,\pm}\\
\psi^2_{n-1,\pm}\\
\psi^3_{n,\pm}\\
\psi^3_{n-1,\pm}
\end{pmatrix}=
\begin{pmatrix}
e^{\pm ik_1}\psi^1_{n,\pm}\\
e^{\pm ik_1}\psi^1_{n-1,\pm}\\
e^{\pm ik_2}\psi^2_{n,\pm}\\
e^{\pm ik_2}\psi^2_{n-1,\pm}\\
e^{\pm ik_3}\psi^3_{n,\pm}\\
e^{\pm ik_3}\psi^3_{n-1,\pm}
\end{pmatrix}\quad , \tilde Y_0\equiv\tilde X'_0\;,\;\tilde X_0"\quad ,
\end{equation}

\noindent
where $\tilde X_{00},\; \tilde X'_0$ and $\tilde X_0"$ denote the transfer
matrices for the leads given by (\ref{14}-\ref{16})
with $\varepsilon_{1n}=\varepsilon_{2n}=\varepsilon_{3n}=0$,
respectively.
The wavenumbers, $k_i$, are defined in terms of the energy $E$
by the eigenvalues of Eqs. (\ref{17}) and (\ref{18}), respectively. By
solving for the eigenvalues we get successively

\begin{eqnarray*}
2\cos k_1
&=&
E-h\quad ,\nonumber\\
2\cos k_2
&=&
E+h\quad , \hspace{6cm}\text{(17.a)}
\end{eqnarray*}

\noindent
for the two-channel system,

\begin{eqnarray*}
2\cos k_1
&=&
E-\sqrt 2 h\quad ,\nonumber\\
2\cos k_2
&=&
E\quad ,\nonumber\\
2\cos k_3
&=&
E+\sqrt 2 h\quad ,\hspace{6cm}\text{(18.a)}
\end{eqnarray*}

\noindent
for the three-channel system with free boundary conditions, whose leads
are
described by $\tilde X'_0$, and, finally,

\begin{eqnarray*}
2\cos k_1
&=&
E-2h\quad ,\nonumber\\
2\cos k_2
&=&
2\cos k_3=E+h\quad , \hspace{5cm}\text{(18.b)}
\end{eqnarray*}

\noindent
for the three-channel model with periodic lateral boundary conditions.
The eigenfunctions in the leads at energy $E$ obtained from
(\ref{12}-\ref{13}) and (\ref{17}-\ref{18}) are of the form

\begin{equation}\label{19}
\psi^{j}_{n,\pm}\sim e^{\pm ink_j}\quad ,
\end{equation}

\noindent
where we choose the wavenumbers $k_j,j=1,2,3$ to be positive, $0\leq
k_j\leq\pi$, so that these functions correspond to plane waves travelling
from left to right and from right to left, respectively.

The transfer matrices for single site slices in Eqs (\ref{12}-\ref{13})
for the leads (i.e.
for $\varepsilon_{in}=0$, $n<1$ or $n>N$) are diagonalized in the bases of
the Bloch plane wave states (\ref{19}).  The diagonalization matrices
which are formed by the eigenvectors of (\ref{17}-\ref{18}), are of the
form

\begin{equation}\label{20}
\tilde V_0=
\begin{pmatrix}
\widehat A_1 & \widehat O\\
\widehat O & \widehat A_2
\end{pmatrix}\quad , \quad
\widehat A_j={1\over\sqrt{2i\sin k_j}}
\begin{pmatrix}
e^{ik_j} & e^{-ik_j}\\
1 & 1
\end{pmatrix}\quad ,
\end{equation}

\noindent
for the two-channel quasi-1D model, and

\begin{equation}\label{21}
\tilde V=
\begin{pmatrix}
\widehat A_1 & \widehat O & \widehat O\\
\widehat O & \widehat A_2 & \widehat O\\
\widehat O & \widehat O & \widehat A_3
\end{pmatrix}\quad ,
\end{equation}

\noindent
with $\widehat A_j$ defined as in (\ref{20}), for the three-channel
models.  The wavenumbers $k_j$ defined by (17a) for the $N=2$ case
and by (18.a) and (18b) for the $N=3$ case with free- and
periodic boundary conditions, respectively.  After finding the inverses
of $\widehat V_0$ and $\widehat V$ and performing the simularity
transformations of $\tilde X_{0n},\;\tilde X'_n$ and $\tilde X_n"$ by
$\widehat V_0$ and $\widehat V$, respectively, we obtain the desired
transfer matrices in the Bloch wave representation of the disordered
wires.  In the two-channel case we find

\begin{eqnarray}\label{22}
\widehat X_{0n}
&\equiv & \widehat V_0^{-1}\tilde X_{0n}\widehat V_0\nonumber \\
&=&
\begin{pmatrix}
e^{ik_1}(1+ia_{1n}) & i\;e^{-ik_1}a_{1n} & -i\;e^{ik_2}b_{n} &
-i\;e^{ik_2}b_{n}\\
-i\;e^{ik_1}a_{1n} & e^{-ik_1}(1-ia_{1n}) & i\;e^{ik_2}b_{n} &
i\;e^{-ik_2}b_{n}\\
-i\;e^{ik_1}b_{n} & -i\;e^{-ik_1}b_{n} & e^{ik_2}(1+ia_{2n}) &
i\;e^{-ik_2}a_{2n}\\
i\;e^{ik_1}b_{n} & i\;e^{-ik_1}b_{n} & -i\;e^{ik_2}b_{n} &
e^{-ik_2}(1-ia_{2n})
\end{pmatrix}\quad ,
\end{eqnarray}

\noindent
where

\begin{eqnarray*}
a_{1n}
&=&
{\varepsilon_{1n}+\varepsilon_{2n}\over 4\sin k_1}\;,\;
a_{2n}{\varepsilon_{1n}+\varepsilon_{2n}\over 4\sin k_2}\quad , \nonumber
\\
b_n
&=&
{\varepsilon_{2n}-\varepsilon_{1n}\over 4\sqrt{\sin k_1\sin k_2}}\quad
,\hspace{8cm}\text{(22.a)}
\end{eqnarray*}

\noindent
are real quantities and $k_1$, $k_2$ are defined by (17a).  For the
three-channel systems we write
the final transfer matrices $\widehat X'_n$ and $\widehat X"_n$, in terms
of a generic matrix

{\footnotesize
\begin{eqnarray}\label{23}
& &
\widehat Z_n=\nonumber \\
& &
\begin{pmatrix}
e^{ik_1}(1+ia_{1n}) & i\;e^{-ik_1}a_{1n} & i\;e^{ik_2}c_n &
i\;e^{-ik_2}c_n & i\;e^{ik_3}g_n & i\;e^{-ik_3}g_n\\
-i\;e^{ik_1}a_{1n} & e^{-ik_1}(1-ia_{1n}) & -i\;e^{ik_2}c_n &
-i\;e^{-ik_2}c_n & -i\;e^{ik_3}g_n & -i\;e^{-ik_3}g_n\\
i\;e^{ik_1}f_n & i\;e^{-ik_1}f_{n} & e^{ik_2}(1+ib_{2n}) &
i\;e^{-ik_2}b_{2n} & i\;e^{ik_3}d_n & i\;e^{-ik_3}d_n\\
-i\;e^{ik_1}f_n & -i\;e^{-ik_1}f_{n} & -i\;e^{ik_2}b_{2n} &
e^{-ik_2}(1-ib_{2n}) & -i\;e^{ik_3}d_n & -i\;e^{-ik_3}d_n\\
i\;e^{ik_1}p_n & i\;e^{-ik_1}p_{n} & i\;e^{ik_2}q_n &
i\;e^{-ik_2}q_n & e^{ik_3}(1+ia_{3n}) & i\;e^{-ik_3}a_{3n}\\
-i\;e^{ik_1}p_n & i\;e^{-ik_1}p_{n} & -i\;e^{ik_2}q_n &
-i\;e^{-ik_2}q_n & -i\;e^{ik_3}a_{3n} & e^{-ik_3}(1-ia_{3n})
\end{pmatrix}\quad ,
\end{eqnarray}}

\noindent
where $a_{1n}, a_{3n}, b_{2n}, c_n, g_n, f_n, d_n, p_n$ and $q_n$ are real
quantities.
In the case of free boundary conditions we obtain $\widehat X'_n=\widehat
V^{-1}\tilde X'_n\widehat V\equiv \widehat Z_n$ with

\begin{eqnarray}\label{24}
a_{1n}
&=&
{\varepsilon_{1n}+2\varepsilon_{2n}+\varepsilon_{3n}\over 8\sin k_1}\;,\;
a_{3n}={\varepsilon_{1n}+2\varepsilon_{2n}+\varepsilon_{3n}\over 8\sin
k_3}\quad ,\nonumber\\
b_{2n}
&=&
{\varepsilon_{1n}+\varepsilon_{3n}\over 4\sin k_2}\;,\;
c_n=f_n={\sqrt 2(\varepsilon_{1n}-\varepsilon_{3n})\over 8\sqrt{\sin
k_1\sin k_2}}\quad ,\nonumber \\
d_n=q_n
&=&
{\sqrt 2(\varepsilon_{1n}-\varepsilon_{3n})\over 8\sqrt{\sin k_2\sin
k_3}}\;,\;
g_n=p_n={\varepsilon_{1n}-2\varepsilon_{2n}+\varepsilon_{3n}\over
8\sqrt{\sin
k_1\sin k_3}}\quad .
\end{eqnarray}

\noindent
Here $k,k_1,k_3$ are defined by (18.a).  On the other hand, for
periodic boundary conditions we find $\widehat X"_n=\widehat V^{-1}\tilde
X"\widehat V\equiv\widehat Z_n$ where

\begin{eqnarray}\label{25}
a_{1n}
&=&
{\varepsilon_{1n}+\varepsilon_{2n}+\varepsilon_{3n}\over 6\sin k_1}\;,\;
a_{3n}={2\varepsilon_{2n}+\varepsilon_{3n}\over 6\sin k_2}\quad
,\nonumber\\
b_{2n}
&=&
{2\varepsilon_{1n}+\varepsilon_{3n}\over 6\sin k_2}\;,\;
c_n={\varepsilon_{1n}-\varepsilon_{3n}\over 6\sqrt{\sin k_1\sin k_2}}\;,\;
g_n={\varepsilon_{2n}-\varepsilon_{3n}\over 6\sqrt{\sin k_1\sin k_2}}\quad
,\nonumber \\
d_n
&=&
{\varepsilon_{3n}-\varepsilon_{2n}\over 6\sin k_2}\;,\;
f_n={2\varepsilon_{1n}-\varepsilon_{2n}-\varepsilon_{3n}\over 6\sqrt{\sin
k_1\sin k_2}}\quad ,\nonumber \\
p_n
&=&
{-\varepsilon_{1n}+2\varepsilon_{2n}-\varepsilon_{3n}\over 6\sqrt{\sin
k_1\sin k_2}}\;,\;
q_n={\varepsilon_{3n}-\varepsilon_{1n}\over 6\sin k_2}\quad ,
\end{eqnarray}

\noindent
where $k_1$ and $k_2=k_3$ are now given by (18b).  Note, in
particular, the diagonalization of the transfer matrices for the leads in
the plane wave bases shown in (\ref{22}-\ref{23}).

Finally, we determine the transfer matrices for the disordered wires of
length $L=N_La$ in terms of the transfer matrices of the individual thin
slices $n$.  As shown by iteration of the transfer equations
(\ref{17}-\ref{18}) rewritten in the Bloch wave basis above, the matrix
transferring an incoming wave at site $n=0$ just outside a disordered wire
to the site $N_L+1$ just beyond its other end is
given by a product of transfer matrices of the form

\begin{equation}\label{26}
\widehat Y_L=\prod^{N_L}_{n=1}\widehat Y_n\quad ,
\end{equation}

\noindent
where $\widehat Y_n$ and $\widehat Y_L$ stand for the three pairs of
transfer matrices $\widehat X_{0n}, \widehat X_{0L}, \widehat X'_n,
\widehat X'_L$ and $\widehat X_n", \widehat X_L"$, respectively, which are
associated with the wire models above\cite{19}.

We shall evaluate the
transfer matrices of the disordered wires for weak disorder to linear
order in the random site energies.  On the other hand, for our explicit
calculations of averages over the disorder below, we assume the site
energies to be independent gaussian random variables with zero mean
values, and correlation

\begin{equation}\label{27}
\langle\varepsilon_{in}\varepsilon_{jm}\rangle=\varepsilon^2_0\delta_{i,j}\delta
_{m,n}\quad .
\end{equation}

\noindent
In this case the site energies corresponding to different slices $n$ in
(\ref{26}) are uncorrelated so that it is indeed sufficient to restrict
the
expansion of the latter expressions to first order in the site energies
for determining averages to order $\varepsilon^2_0$.
The
transfer matrices $\widehat Y_n$ given by (\ref{22}-\ref{23}) are sums,

\begin{equation}\label{28}
\widehat Y_n=\widehat Y^{(0)}+\widehat Y_n^{(1)}\quad ,
\end{equation}

\noindent
of a zeroth order diagonal matrix $\widehat Y^{(0)}$ independent of the
site energies (transfer matrix of the leads) and a matrix which is linear
in the energies $\varepsilon_{jn}$.  By inserting (\ref{28}) into
(\ref{26})
we obtain to first order

\begin{equation}\label{29}
\widehat Y_L=(\widehat Y^{(0)})^{N_L}+\sum^{N_L}_{m=1}(\widehat
Y^{(0)})^{m-1}
\widehat Y^{(1)}_m(\widehat Y^{(0)})^{N_L-m}+\cdots\quad .
\end{equation}

\noindent
Next we insert the slice matrices $\widehat Y^{(0)}$ and $\widehat
Y^{(1)}_m$ from (\ref{22}-\ref{23}) for our various quasi-1D systems and
obtain successively:

\begin{eqnarray}\label{30}
& &
\widehat
X_{0L}=\text{diag}(e^{ik_1N_L},e^{-ik_1N_L},e^{ik_2N_L},e^{-ik_2N_L})\nonumber \\
& &
+\sum^{N_L}_{m=1}
\begin{pmatrix}
ia_{1m}e^{ik_1}u_1 & ia_{1m}e^{-ik_1}s^*_1 & -ib_{m}e^{ik_2}w_{21} &
-ib_{1m}e^{-ik_2}v^*_{21}\\
-ia_{1m}e^{ik_1}s_1 & -ia_{1m}e^{-ik_1}u^*_1 & ib_{m}e^{ik_2}v_{21} &
ib_{1m}e^{-ik_2}w^*_{21}\\
-ib_{m}e^{ik_1}w_{12} & -ib_{m}e^{-ik_1}v^*_{12} & ia_{2m}e^{ik_2}u_{2} &
ia_{2m}e^{-ik_2}s^*_{2}\\
ib_{m}e^{ik_1}v_{12} & ib_{m}e^{-ik_1}w^*_{12} & -ia_{2m}e^{ik_2}s_{2} &
-ia_{2m}e^{-ik_2}u^*_{2}\\
\end{pmatrix}\quad ,
\end{eqnarray}

\noindent
for the two-channel case.  Here

\begin{eqnarray}\label{31}
s_j
&=&
e^{i(N_L-2m+1)k_j}\;,\;
u_j=e^{i(N_L-1)k_j}\quad ,\nonumber\\
v_{ij}
&=&
e^{i(N_L-m)k_i-i(m-1)k_j}\;,\;w_{ij}=e^{i(N_L-m)k_i+i(m-1)k_j}\quad ,
\end{eqnarray}

\noindent
where $k_1$ and $k_2$ are defined by (17a);
{\footnotesize
\begin{eqnarray}\label{32}
\widehat
X'_{L}&=&\text{diag}\;(e^{ik_1N_L},e^{-ik_1N_L},e^{ik_2N_L},e^{-ik_2N_L},e^{ik_3
N_L},e^{-ik_3N_L})
+
\sum^{N_L}_{m=1}\nonumber\\
& &
\begin{pmatrix}
ia_{1m}e^{ik_1}u_1 & ia_{1m}e^{-ik_1}s^*_1 & ic_{m}e^{ik_2}w_{21} &
ic_{m}e^{-ik_2}v^*_{21} & ig_{m}e^{ik_3}w_{31} &
ig_{m}e^{-ik_3}v^*_{31}\\
-ia_{1m}e^{ik_1}s_1 & -ia_{1m}e^{-ik_1}u^*_1 & -ic_{m}e^{ik_2}v_{21} &
-ic_{m}e^{-ik_2}w^*_{21} & -ig_{m}e^{ik_3}v_{31} &
-ig_{m}e^{-ik_3}w^*_{31}\\
if_{m}e^{ik_1}w_{12} & if_{m}e^{-ik_1}v^*_{12} & ib_{2m}e^{ik_2}u_{2} &
ib_{2m}e^{-ik_2}s^*_{2} & id_{m}e^{ik_3}w_{32} & id_{m}e^{-ik_3}v^*_{32}\\
-if_{m}e^{ik_1}v_{12} & -if_{m}e^{-ik_1}w^*_{12} & -ib_{2m}e^{ik_2}s_{2} &
-ib_{2m}e^{-ik_2}u^*_{2} & -id_{m}e^{ik_3}v_{32} &
-id_{m}e^{-ik_3}w^*_{32}\\
ip_{m}e^{ik_1}w_{13} & ip_{m}e^{-ik_1}v^*_{13} & iq_{m}e^{ik_2}w_{23} &
iq_{m}e^{-ik_2}v^*_{23} & ia_{3m}e^{ik_3}u^*_{3} &
ia_{3m}e^{-ik_3}s^*_{3}\\
-ip_{m}e^{ik_1}v_{13} & -ip_{m}e^{-ik_1}w^*_{13} & -iq_{m}e^{ik_2}w_{23} &
-iq_{m}e^{-ik_2}w^*_{23} & -ia_{3m}e^{ik_3}s^*_{3} &
-ia_{3m}e^{-ik_3}u^*_{3}\\
\end{pmatrix}\quad ,
\end{eqnarray}}

\noindent
for the three-channel case with free boundary conditions where
$k_1,k_2,k_3$
in the definitions (\ref{24}) and (\ref{31}) are given by (18.a); the
transfer matrix $\widehat X_L"$ for the three-channel model with periodic
boundary conditions is given by (\ref{32}), using the definitions
(\ref{25}) of the slice parameters $a_{1m}, a_{3m}, b_{2m}, c_m$ and the
definition (18b) of the wavenumbers $k_1, k_2, k_3$.

\subsection{Scattering matrices}

The scattering of plane waves (reflection and transmission) at and between
the two ends of the random quasi-1D systems is governed by the $S$-matrix,

\begin{equation}\label{33}
\widehat S=
\begin{pmatrix}
\hat r^{-+} & \hat t^{--}\\
\hat t^{++} & \hat r^{+-}
\end{pmatrix}\quad ,
\end{equation}

\noindent
where

\begin{equation}\label{34}
\hat t^{\mp\mp}=
\begin{pmatrix}
 t^{\mp\mp}_{11} &  t^{\mp\mp}_{12} & \cdots\\
 t^{\mp\mp}_{21} &  t^{\mp\mp}_{22} & \cdots\\
\vdots & \vdots & \vdots
\end{pmatrix}\quad ,
\end{equation}

\noindent
and

\begin{equation}\label{35}
\hat r^{\pm\mp}=
\begin{pmatrix}
 r^{\pm\mp}_{11} &  r^{\pm\mp}_{12} & \cdots\\
 r^{\pm\mp}_{21} &  r^{\pm\mp}_{22} & \cdots\\
\vdots & \vdots & \vdots
\end{pmatrix}\quad .
\end{equation}

\noindent
Here $t_{ij}^{++} (t_{ij}^{--})$ and $r_{ij}^{-+}(r_{ij}^{+-})$ denote the
transmitted and reflected amplitudes in channel $i$ when there is a unit
flux incident from the left (right) in channel $j$.  Left to right- and
right to left directions are denoted by + and -, respectively.  The
$S$-matrix expresses outgoing wave amplitudes in terms of ingoing ones on
either side of the quasi-1D disordered wire via the scattering
relations\cite{11,21}

\begin{equation}\label{36}
\begin{pmatrix}
0 \\ 0'
\end{pmatrix}=
\widehat S
\begin{pmatrix}
I \\ I'
\end{pmatrix}\quad .
\end{equation}

\noindent
Here $I$ and $I'$ (0 and $0'$) denote ingoing (outgoing) amplitudes at the
left and right sides of the disordered region, respectively.  It follows
from current conservation that e.g. for a unit flux which is incident from
the right in channel $i$ one has

\begin{equation}\label{37}
\sum^{N}_{j=1}(\mid t^{--}_{ji}\mid^2+\mid r^{-+}_{ji}\mid^2)=1\quad .
\end{equation}

\noindent
Likewise, one has also

\begin{equation*}
\sum^{N}_{j=1}(\mid t^{++}_{ji}\mid^2+\mid r^{+-}_{ji}\mid^2)=1\quad
.\hspace{6cm}\text{(37.a)}
\end{equation*}

Our task is now to derive miscroscopic  realizations of the $S$-matrix in
terms of the transfer matrices (\ref{30}) and (\ref{32}) describing
transfer of Bloch waves across finite quasi-1D disordered systems.  Let us
remark that this may not always be possible as is seen here in
the case of the three channel periodic model.  Indeed, in this case we
are able to identify proper transmission and reflection amplitudes obeying
the symmetry relations (\ref{37}) and (37a) only when assuming the
random site energies on chains 1 and 2 to be identical,

\begin{equation}\label{38}
\varepsilon_{1n}=\varepsilon_{2n}\;,\;n=1,2,\ldots N_L\quad ,
\end{equation}

\noindent
rather than allowing the energies at all pairs of sites of the quasi-1D
system to be uncorrelated, as in (\ref{27}).  In the absence of the
correlation (\ref{38}) the obtained transport amplitudes cannot be
identified as
actual reflection and transmission amplitudes of Bloch waves.

If the amplitude at the $n$th site in a channel $j$ corresponds to a Bloch
wave $\psi_n^j=e^{ink_j}$ (equation (\ref{19})) then the $j,n$ and $j,n-1$
components of wave amplitude vectors,

$$\widehat W^{-1}
\begin{pmatrix}
\vdots \\ \psi^j_n \\ \psi^j_{n-1} \\ \vdots
\end{pmatrix}$$

\noindent
(with $\widehat W\equiv\widehat V_0$ or $\widehat V$), being transferred
by
the $n$-th slice have values $a\;e^{inkj}$ and 0, respectively; on the
other hand if $\psi^j_n=e^{-inkj}$ then the $j,n$ and $j,n-1$ components
of the above vectors are 0 and $a'e^{-inkj}$.  Hence, in accordance with
our notation above for transmission and reflection matrix elements we
denote the amplitudes at site $j,n$ and $j,n-1$ respectively as
$a^+_{j,n-1}$
and $a^-_{j,n-1}$ since they correspond to amplitudes being transferred by the
$n$-th slice and propagating in the $k_j$ and $-k_j$ directions,
respectively.  The transformed amplitude vectors

$$\widehat W^{-1}\begin{pmatrix}
\vdots\\\psi^j_n\\\psi^j_{n-1}\\\vdots
\end{pmatrix}$$

\noindent and

$$\widehat W^{-1}\begin{pmatrix}
\vdots\\\psi^j_{n+1}\\\psi^j_n\\\vdots
\end{pmatrix}$$

\noindent

are thus rewritten, respectively, as

\begin{equation}\label{39}
\widehat W^{-1}
\begin{pmatrix}
\psi^1_{n} \\ \psi^1_{n-1}\\ \psi^2_{n} \\ \psi^2_{n-1} \\\vdots
\end{pmatrix}
\equiv
\begin{pmatrix}
a^{+}_{1,n-1} \\ a^{-}_{1,n-1} \\ a^{+}_{2,n-1} \\ a^{-}_{2,n-1} \\\vdots
\end{pmatrix}
\quad \text{and}\quad
\widehat W^{-1}
\begin{pmatrix}
\psi^1_{n+1} \\ \psi^1_{n}\\ \psi^2_{n+1} \\ \psi^2_{n} \\\vdots
\end{pmatrix}
\equiv
\begin{pmatrix}
a^{+}_{1,n} \\ a^{-}_{1,n} \\ a^{+}_{2,n} \\ a^{-}_{2,n} \\\vdots
\end{pmatrix}\quad .
\end{equation}

\noindent
Using a similar notation for wave amplitudes transferred from $n=0$ to
$n=N_L$ across a disordered wire of length $L=N_La$, the wave transfer
equations in the Bloch representation, obtained by iterating
(\ref{12}-\ref{13}) read

\begin{equation}\label{40}
\begin{pmatrix}
a^{+}_{1,L} \\ a^{-}_{1,L} \\ a^{+}_{2,L} \\ a^{-}_{2,L}
\end{pmatrix}=
\widehat X_{0L}
\begin{pmatrix}
a^{+}_{1,0} \\ a^{-}_{1,0} \\ a^{+}_{2,0} \\ a^{-}_{2,0}
\end{pmatrix}\quad ,
\end{equation}

\noindent
and

\begin{equation}\label{41}
\begin{pmatrix}
a^{+}_{1,L} \\ a^{-}_{1,L} \\ a^{+}_{2,L} \\ a^{-}_{2,L} \\  a^{+}_{3,L}
\\
a^{-}_{3,L}
\end{pmatrix}=
\widehat Y_L
\begin{pmatrix}
a^{+}_{1,0} \\ a^{-}_{1,0} \\ a^{+}_{2,0} \\ a^{-}_{2,0} \\ a^{+}_{3,0} \\
a^{-}_{3,0}
\end{pmatrix}\quad ,
\end{equation}

\noindent
where $\widehat X_{0L}$ is given in (\ref{30}) and $\widehat Y_L$ stands
for $\widehat X'_L$ and $\widehat X"_L$ in (\ref{32}) with parameters
defined by (18.a) and (\ref{24}) and by (18.b) and (\ref{25}),
respectively.

In order to derive the $S$-matrices for our two- and three channel wire
models we first rewrite (\ref{40}-\ref{41}) in the forms of equations
involving outgoing amplitudes on the left side and incoming ones as the
right side, as in (\ref{36}).  In the notation of (\ref{40})
and (\ref{41}) we have e.g. for $N=3$

\begin{equation}\label{42}
(I)\equiv
\begin{pmatrix}
a^{+}_{1,0} \\ a^{+}_{2,0}\\ a^{+}_{3,0}
\end{pmatrix}\; ,\;
(I')\equiv
\begin{pmatrix}
a^{-}_{1,L} \\ a^{-}_{2,L}\\ a^{-}_{3,L}
\end{pmatrix}\; ,\;
(O)\equiv
\begin{pmatrix}
a^{-}_{1,0} \\ a^{-}_{2,0}\\ a^{-}_{3,0}
\end{pmatrix}\; ,\;
(O')\equiv
\begin{pmatrix}
a^{+}_{1,L} \\ a^{+}_{2,L}\\ a^{+}_{3,L}
\end{pmatrix}\quad ,
\end{equation}

\noindent
and so we rearrange (\ref{40}-\ref{41}) in the form

\begin{equation}\label{43}
\widehat A_{2,3}
\begin{pmatrix}
O \\ O'
\end{pmatrix}=
\widehat B_{2,3}
\begin{pmatrix}
I \\ I'
\end{pmatrix}\quad ,
\end{equation}

\noindent
where the pairs of matrices $\widehat A_2$ and $\widehat B_2$ and
$\widehat
A_3$ and $\widehat B_3$ correspond to two- and three channel cases,
respectively.  Using the notation $(\widehat X_{0L})_{ij}\equiv X_{ij}$
and $(\widehat Y_L)_{ij}\equiv Y_{ij}$ for the matrix elements in
(\ref{30})
and (\ref{32}), respectively, we find

\begin{equation}\label{44}
\widehat A_2=
\begin{pmatrix}
-X_{12} & -X_{14} & 1 & 0\\
-X_{22} & -X_{24} & 0 & 0\\
-X_{32} & -X_{34} & 0 & 1\\
-X_{42} & -X_{44} & 0 & 0
\end{pmatrix}\; ,\;
\widehat B_2=
\begin{pmatrix}
X_{11} & X_{13} & 0 & 0\\
X_{21} & X_{23} & -1 & 0\\
X_{31} & X_{33} & 0 & 0\\
X_{41} & X_{43} & 0 & -1
\end{pmatrix}\quad ,
\end{equation}

\begin{equation}\label{45}
\widehat A_3=
\begin{pmatrix}
-Y_{12} & -Y_{14} & -Y_{16} &1 & 0 & 0\\
-Y_{22} & -Y_{24} & -Y_{26} &0 & 0 & 0\\
-Y_{32} & -Y_{34} & -Y_{36} &0 & 1 & 0\\
-Y_{42} & -Y_{44} & -Y_{46} &0 & 0 & 0\\
-Y_{52} & -Y_{54} & -Y_{56} &0 & 0 & 1\\
-Y_{62} & -Y_{64} & -Y_{66} &0 & 0 & 0
\end{pmatrix}\; ,\;
\widehat B_3=
\begin{pmatrix}
Y_{11} & Y_{13} & Y_{15} &0 & 0 & 0\\
Y_{21} & Y_{23} & Y_{25} &-1 & 0 & 0\\
Y_{31} & Y_{33} & Y_{35} &0 & 0 & 0\\
Y_{41} & Y_{43} & Y_{45} &0 & -1 & 0\\
Y_{51} & Y_{53} & Y_{55} &0 & 0 & 0\\
Y_{61} & Y_{63} & Y_{65} &0 & 0 & -1
\end{pmatrix}\quad .
\end{equation}

\noindent
The $S$-matrices for the two- and three channel cases are those given by
$\widehat S=\widehat A^{-1}_2\widehat B_2$ and $\widehat S=\widehat
A^{-1}_3\widehat B_3$, respectively.  After inverting $\widehat A_2$ and
$\widehat A_3$ we finally obtain, successively for the two- and three
channel cases:

\begin{equation}\label{46}
\widehat S={1\over \delta}
\begin{pmatrix}
\delta_1 & \delta_2 & X_{44} & -X_{24}\\
\delta_3 & \delta_4 & -X_{42} & X_{22}\\
X_{11}\delta+X_{21}\delta_5 & X_{13}\delta+X_{23}\delta_5 & -\delta_5 &
-\delta_6\\
+X_{41}\delta_6 & +X_{43}\delta_6 & & \\
X_{31}\delta+X_{21}\delta_7 & X_{33}\delta+X_{23}\delta_7 & -\delta_7 &
-\delta_8\\
+X_{41}\delta_8 & +X_{43}\delta_8 & & \\
\end{pmatrix}\quad ,
\end{equation}

\noindent
where

\begin{eqnarray}\label{47}
\delta
&=&
X_{22}X_{44}-X_{24}X_{42}\;,\;
\delta_1=X_{24}X_{41}-X_{44}X_{21}\;,\;
\delta_2=X_{24}X_{43}-X_{44}X_{23}\quad ,\nonumber \\
\delta_3
&=&
X_{42}X_{21}-X_{41}X_{22}\;,\;
\delta_4=X_{42}X_{23}-X_{22}X_{43}\;,\;
\delta_5=X_{42}X_{14}-X_{12}X_{44}\quad ,\nonumber \\
\delta_6
&=&
X_{12}X_{24}-X_{22}X_{14}\;,\;
\delta_7=X_{42}X_{34}-X_{32}X_{44}\;,\;
\delta_8=X_{32}X_{24}-X_{22}X_{34}\quad ,
\end{eqnarray}

\noindent
are second order subdeterminants of $\widehat X_{0L}$;

\begin{equation}\label{48}
\widehat S=
\begin{pmatrix}
\widehat S_1 & \widehat S_3\\
\widehat S_2 & \widehat S_4
\end{pmatrix}\quad ,
\end{equation}

\noindent
where

\begin{eqnarray*}
& &
\widehat S_1={1\over\Delta}\\
& &
\begin{pmatrix}
-\beta_{1}Y_{21}-\beta_{4}Y_{41}-\beta_{7}Y_{61}&
-\beta_{1}Y_{23}-\beta_{4}Y_{43}-\beta_{7}Y_{63}&
-\beta_{1}Y_{25}-\beta_{4}Y_{45}-\beta_{7}Y_{65}\\
\beta_{2}Y_{21}+\beta_{5}Y_{41}+\beta_{8}Y_{61}&
\beta_{2}Y_{23}+\beta_{5}Y_{43}+\beta_{8}Y_{63}&
\beta_{2}Y_{25}+\beta_{5}Y_{45}+\beta_{8}Y_{65}\\
-\beta_{3}Y_{21}-\beta_{6}Y_{41}-\beta_{9}Y_{61}&
-\beta_{3}Y_{23}-\beta_{6}Y_{43}-\beta_{9}Y_{63}&
-\beta_{3}Y_{25}-\beta_{6}Y_{45}-\beta_{9}Y_{65}
\end{pmatrix}\quad ,\\
& & \hspace{16cm}\text{(48.a)}
\end{eqnarray*}

{\footnotesize
\begin{eqnarray*}
& &
\widehat S_2={1\over\Delta}\\
& &
\begin{pmatrix}
Y_{11}\Delta+Y_{21}\Delta_1+Y_{41}\Delta_4+Y_{61}\Delta_7 \quad &
Y_{13}\Delta+Y_{23}\Delta_1+Y_{43}\Delta_4+Y_{63}\Delta_7 \quad &
Y_{15}\Delta+Y_{25}\Delta_1+Y_{45}\Delta_4+Y_{65}\Delta_7 \\
Y_{31}\Delta-Y_{21}\Delta_2-Y_{41}\Delta_5-Y_{61}\Delta_8 \quad &
Y_{33}\Delta-Y_{23}\Delta_2-Y_{43}\Delta_5-Y_{63}\Delta_8 \quad &
Y_{35}\Delta-Y_{25}\Delta_2-Y_{45}\Delta_5-Y_{65}\Delta_8\\
Y_{51}\Delta+Y_{21}\Delta_3+Y_{41}\Delta_6+Y_{61}\Delta_9 \quad &
Y_{53}\Delta+Y_{23}\Delta_3+Y_{43}\Delta_6+Y_{63}\Delta_9 \quad &
Y_{55}\Delta+Y_{25}\Delta_3+Y_{45}\Delta_6+Y_{65}\Delta_9
\end{pmatrix}\quad ,\\
& & \hspace{16cm}\text{(48.b)}
\end{eqnarray*}}

\begin{equation*}
\widehat S_3={1\over\Delta}
\begin{pmatrix}
\beta_1 & \beta_4 & \beta_7\\
-\beta_2 & -\beta_5 & -\beta_8\\
\beta_3 & \beta_6 & \beta_9
\end{pmatrix}\quad ,\hspace{6cm}\text{(48.c)}
\end{equation*}

\begin{equation*}
\widehat S_4={1\over\Delta}
\begin{pmatrix}
-\Delta_1 & -\Delta_4 & -\Delta_7\\
\Delta_2 & \Delta_5 & \Delta_8\\
-\Delta_3 & -\Delta_6 & -\Delta_9
\end{pmatrix}\quad ,\hspace{6cm}\text{(48.d)}
\end{equation*}

\noindent
which involve subdeterminants of second order of $\widehat Y_L$,

\begin{eqnarray*}
\beta_1
&=&
Y_{46}Y_{64}-Y_{44}Y_{66}\;,\;
\beta_2=Y_{62}Y_{46}-Y_{42}Y_{66}\;,\;
\beta_3=Y_{62}Y_{44}-Y_{42}Y_{64}\quad, \\
\beta_4
&=&
Y_{24}Y_{66}-Y_{64}Y_{26}\;,\;
\beta_5=Y_{22}Y_{66}-Y_{62}Y_{26}\;,\;
\beta_6=Y_{22}Y_{64}-Y_{62}Y_{24}\quad, \\
\beta_7
&=&
Y_{44}Y_{26}-Y_{24}Y_{46}\;,\;
\beta_8=Y_{42}Y_{26}-Y_{22}Y_{46}\;,\;
\beta_9=Y_{24}Y_{42}-Y_{22}Y_{44}\quad, \hspace{2cm}\text{(48.e)}
\end{eqnarray*}

\noindent
as well as third order subderterminants of $\widehat A_3$ which result
from the minors of various elements,

\begin{eqnarray*}
\Delta_1
&=&
\text{min} (\widehat A_3)_{24}\;,\;
\Delta_2=\text{min}(\widehat A_3)_{25}\quad, \\
\Delta_3
&=&
\text{min} (\widehat A_3)_{26}\;,\;
\Delta_4=\text{min}(\widehat A_3)_{44}\;,\;
\Delta_5=\text{min}(\widehat A_3)_{45}\quad, \\
\Delta_6
&=&
\text{min} (\widehat A_3)_{46}\;,\;
\Delta_7=\text{min}(\widehat A_3)_{64}\;,\;
\Delta_8=\text{min}(\widehat A_3)_{65}\quad, \\
\Delta_9
&=&
\text{min}(\widehat A_3)_{66}\;,\text{and}\;
\Delta=\text{det}\;\widehat A_3\quad. \hspace{6cm}\text{(48.f)}
\end{eqnarray*}

\noindent
The partial matrices $\widehat S_j,j=1,2,3,4$ in (\ref{48}) clearly
correspond to reflection and transmission matrices in (\ref{33}).

The matrices (\ref{46}) and (\ref{48},\ref{48}a-d) are the
$S$-matrix expressions for the two- and three channel quasi-1D systems in
terms of
characteristic quantum channel wavenumbers and of the tight-binding
quantities in the transfer matrices (\ref{30}) and (\ref{32}) for weak
disorder.  Identification of these expressions with (\ref{33}-\ref{35})
yields the transmission- anf reflection matrices for these quasi-1D
disordered models under the proviso that the symmetry relation
(\ref{37}-37a) are obeyed.

\section{RESULTS AND DISCUSSION}

The results for the transmission- and reflection matrices of two- and
three channel tight-binding wires are applied in this section for finding
the averaged transmission and reflection coefficients associated with the
various channels, for weak disorder.  These results allow us, in
particular, to explicitely check the symmetry property (\ref{37}) in the
two-channel case, as well as in the three-channel case with free boundary
conditions.  On the other hand in the three-channel case with periodic
boundary conditions we show that (\ref{37}) is obeyed if one restricts the
disorder to a correlated site energy disorder with identical site energies
(\ref{38}) on chains 1 and 2 and independent random energies on chain 3.

The results for the averaged transmission coefficients are used for
obtaining the length of exponential localization from (\ref{11}) and
(\ref{1}).  Our exact quantum expressions of localization lengths for weak
disorder reduce to the well-known Thouless expression for a 1D chain in
the
limit of vanishing interchain coupling ($h\rightarrow 0$).  It is useful,
before presenting our results, to briefly recall the derivation of
Thouless' result from the transfer matrix approach\cite{19}.  For a single
disordered chain of length $L$ the localization length is given by

\begin{equation}\label{50}
{1\over L_c}=-\lim_{N_L\rightarrow\infty}(2N_l)^{-1}\langle\ln\mid
t\mid^2\rangle\quad ,
\end{equation}

\noindent
where $\mid t\mid^2=\mid t^{--}\mid^2=\mid t^{++}\mid^2$ is the
transmission coefficient
which is related to the
two-dimensional transfer matrices $\widehat X_n$ (in the Bloch wave
representation) for thin sections enclosing the $n$th site by\cite{19}

\begin{equation}\label{51}
{1\over t^{--}}=\left(\prod^{N_L}_{n=1}\widehat X_n\right)_{22}\quad .
\end{equation}

\noindent
This relation follows by transforming the $S$-matrix for scattering states
into a transfer matrix whose 22-element is $1/t^{--}$.  Expanding the
transfer matrix for the whole chain, $\prod^{N_L}_{n=1}\widehat X_n$, to
first order in the uncorrelated site energies and performing the average
in (\ref{50}), using (\ref{27}), yields

\begin{equation}\label{52}
{1\over L_c}\equiv{1\over\xi}={\varepsilon^2_0\over 8\sin^2 k}\;,\;
E=2\cos k\quad ,
\end{equation}

\noindent
which is Thouless' expression for the localization length in
the
tight-binding band, for weak gaussian disorder\cite{13}.

For convenience of the following discussion for two- and three channel
systems, the explicit forms of the transmission and reflection
coefficients
obtained by identifying the $S$-matrix (\ref{33}-\ref{35}) successively
with (\ref{46}) and (\ref{48}, 48.a-d) and replacing the transfer
matrix elements entering in these expressions by their explicit forms in
(\ref{30}-\ref{32}) are given in the appendix.

\subsection*{A. Two-channel wires}

By averaging the partial transmission- and reflection coefficients given
by (A.1-A.4) over the disorder, using (22.a) and (\ref{27}), we
obtain successively

\begin{equation}\label{53}
\langle\mid t_{11}^{--}\mid^2\rangle=1-{N_L\varepsilon^2_0\over 8}
\left({1\over\sin^2 k_1}+{2\over\sin k_1\sin k_2}\right)\quad ,
\end{equation}

\begin{equation}\label{54}
\langle\mid t_{22}^{--}\mid^2\rangle=1-{N_L\varepsilon^2_0\over 8}
\left({1\over\sin^2 k_2}+{2\over\sin k_1\sin k_2}\right)\quad ,
\end{equation}

\begin{equation}\label{55}
\langle\mid t_{12}^{--}\mid^2\rangle=\langle\mid
t_{21}^{--}\mid^2\rangle={N_L\varepsilon^2_0\over 8\sin k_1\sin k_2}\quad ,
\end{equation}

\begin{equation}\label{56}
\langle\mid r_{11}^{-+}\mid^2\rangle={N_L\varepsilon^2_0\over 8\sin^2
k_1}\quad ,
\end{equation}

\begin{equation}\label{57}
\langle\mid r_{22}^{-+}\mid^2\rangle={N_L\varepsilon^2_0\over 8\sin^2
k_2}\quad ,
\end{equation}

\begin{equation}\label{58}
\langle\mid r_{12}^{-+}\mid^2\rangle=\langle\mid
r_{21}^{-+}\mid^2\rangle{N_L\varepsilon^2_0\over 8\sin k_1\sin
k_2}\quad .
\end{equation}

\noindent
The expressions (\ref{53}-\ref{58}) are consistent with the symmetry
relations (\ref{37}) resulting from current conservation.

The inverse localization length for weak disorder is obtained by expanding
(\ref{11}) to lowest order in the random site energies using (\ref{1}) and
(\ref{53}-\ref{55}).  It is given by

\begin{equation}\label{59}
{1\over L_c}\equiv{1\over L_{0c}}={\varepsilon^2_0\over 32}
\left({1\over\sin k_1}+{1\over\sin k_2}\right)^2\quad ,
\end{equation}

\noindent
for energies $E$ restricted to the Bloch bands $E=h+2\cos k_1$, and
$E=-h+2\cos k_2$.  This exact expression for weak disorder reveals three
important properties:

\begin{enumerate}
\item It proves miscoscopically that all states in the Bloch energy bands
of two-channel quasi-1D disordered systems are localized.

\item In the absence of interchain hopping $(h=0)$ it reduces to the
localization length (\ref{52}) for a 1D chain described by the Anderson
model.  For weak interchain hopping (\ref{59}) becomes

\begin{equation}\label{60}
{1\over L_{0c}}={\varepsilon^2_0\over 8\sin^2 k}
\left[1+{h^2\over 4\sin^2k}(1+3\cot^2k)+O(h^4)\right]\quad ,
\end{equation}

\noindent
which is valid for energies sufficiently close to the band centre of the
$h=0$ energy band, $E=2\cos k$.  This shows that a weak interchain hopping
enhances localization in comparison to the purely 1D case, i.e.
$L_{0c}<\xi$.

\item For large interchain hopping rates, i.e. $\mid h\mid>>\mid E\mid$
(where $E$ is of the order of the fermi energy) we have $\sin
k_{1,2}\simeq\sqrt{1-h^2/4}\;,\;\mid h\mid/2\leq 1$, which yields

\begin{equation}\label{61}
{1\over L_c}\simeq{\varepsilon^2_0\over 2}{1\over 4-h^2}={1\over
\xi_0}{4\over 4-h^2}\end{equation}

\noindent
where $\xi_0=\varepsilon^2_0/8$ is the 1D localization length (\ref{52}) at
the band centre.  Thus at large hopping rates (with $\mid h\mid<2$)
localization is also enhanced in comparison to 1D localization.
\end{enumerate}

\subsection*{B. Three-channel wires}

\begin{enumerate}
\item {\em Free boundary conditions}

\qquad The evaluation of the disorder averages of the transmission- and
reflection coefficients in (A.5-A.6), using (\ref{27}) and the explicit
expressions in (A.7-A.9) with the tight-binding parameters (\ref{24})
yields, to order $\varepsilon^2_0$,

\begin{equation}\label{62}
\langle\mid t_{11}^{--}\mid^2\rangle=
1-{N_L\varepsilon^2_0\over 32}\left({3\over\sin^2k_1}+{6\over\sin k_1\sin
k_3}+{4\over\sin k_1\sin k_2}\right)\quad ,
\end{equation}

\begin{equation}\label{63}
\langle\mid t_{22}^{--}\mid^2\rangle=
1-{N_L\varepsilon^2_0\over 32}\left({4\over\sin^2k_2}+{4\over\sin k_2\sin
k_3}+{4\over\sin k_1\sin k_2}\right)\quad ,
\end{equation}

\begin{equation}\label{64}
\langle\mid t_{33}^{--}\mid^2\rangle=
1-{N_L\varepsilon^2_0\over 32}\left({3\over\sin^2k_3}+{6\over\sin k_1\sin
k_3}+{4\over\sin k_2\sin k_3}\right)\quad ,
\end{equation}

\begin{equation}\label{65}
\langle\mid t_{12}^{--}\mid^2\rangle=\langle\mid t_{21}^{--}\mid^2\rangle
={2N_L\varepsilon^2_0\over 32\sin k_1\sin k_2}\quad ,
\end{equation}

\begin{equation}\label{66}
\langle\mid t_{13}^{--}\mid^2\rangle=\langle\mid t_{31}^{--}\mid^2\rangle
={3N_L\varepsilon^2_0\over 32\sin k_1\sin k_3}\quad ,
\end{equation}

\begin{equation}\label{67}
\langle\mid t_{23}^{--}\mid^2\rangle=\langle\mid t_{32}^{--}\mid^2\rangle
={2N_L\varepsilon^2_0\over 32\sin k_2\sin k_3}\quad ,
\end{equation}

\begin{equation}\label{68}
\langle\mid r_{11}^{-+}\mid^2\rangle={3N_L\varepsilon^2_0\over
32\sin^2k_1}\quad ,
\end{equation}

\begin{equation}\label{69}
\langle\mid r_{22}^{-+}\mid^2\rangle={4N_L\varepsilon^2_0\over
32\sin^2k_2}\quad ,
\end{equation}

\begin{equation}\label{70}
\langle\mid r_{33}^{-+}\mid^2\rangle={3N_L\varepsilon^2_0\over
32\sin^2k_3}\quad ,
\end{equation}

\begin{equation}\label{71}
\langle\mid r_{12}^{-+}\mid^2\rangle=\langle\mid
r_{21}^{-+}\mid^2\rangle{2N_L\varepsilon^2_0\over \sin k_1\sin
k_2}\quad ,
\end{equation}

\begin{equation}\label{72}
\langle\mid r_{13}^{-+}\mid^2\rangle=\langle\mid
r_{31}^{-+}\mid^2\rangle{3N_L\varepsilon^2_0\over \sin k_1\sin
k_3}\quad ,
\end{equation}

\begin{equation}\label{73}
\langle\mid r_{23}^{-+}\mid^2\rangle=\langle\mid
r_{32}^{-+}\mid^2\rangle{2N_L\varepsilon^2_0\over \sin k_2\sin
k_3}\quad .
\end{equation}

\noindent
Again, the expressions (\ref{62}-\ref{73}) obey the current conservation
property (\ref{37}) for the three channel case: the reduction of the
intra-channel transmission coefficients due to scattering by the
disordered wire is exactly compensated by the occurence of interchannel
transmissions and by reflections.

\qquad For the inverse localization length we obtain, from
(\ref{11}),(\ref{1})
and (\ref{62}-\ref{67}),

\begin{equation}\label{74}
{1\over L_c}={\varepsilon^2_0\over 64}
\left({1\over \sin^2k_1}+{4\over
3}{1\over\sin^2k_2}+{1\over\sin^2k_3}+{4\over3\sin k_1\sin k_2}+{4\over3\sin
k_2\sin k_3}+{2\over\sin k_1\sin k_3}\right)\quad ,
\end{equation}

\noindent
where $k_1,k_2,k_3$ are defined by (18.a).  Like (\ref{59}), this
expression is exact to order $\varepsilon^2_0$ for weak disorder.  It
demonstrates that the eigenstates in the Bloch bands of the three channel
quasi-1D system with free boundary conditions are localized.  It too
reduces to the Thouless result (\ref{52}) in 1D for vanishing interchain
hopping.  For small values of $\mid h\mid$, in particular for $\mid
h\mid <<1$ for $E\rightarrow 0$, (\ref{74}) reduces to

\begin{equation}\label{75}
{1\over L_c}={\varepsilon^2_0\over 8\sin^2k_2}
\left[1+{h^2\over\sin^2k_2}\left(1+{1\over 2}\cos^2k_2+{1\over
4}\cot^2k_2\right)+O(h^4)\right]\quad ,
\end{equation}

\noindent
and for large values $(h^2>>E^2/2)$ restricted to $\mid h\mid<2$ it becomes

\begin{equation}\label{76}
{1\over L_c}\simeq{\varepsilon^2_0\over 8}{1\over 2-h^2}\quad ,
\end{equation}

\noindent
which shows, in particular, that $L_c$ is increased with respect to the 1D
value at the band centre ($\xi_0$) for $\mid h\mid$ (in units of the
interchain hopping rate) less than 1.  Thus, the domain of
interchain hopping rates in which (\ref{74}) leads to localization lengths
larger than $\xi_0$ is defined by

\begin{equation}\label{77}
\mid E\mid<<\sqrt 2 \mid h\mid<\sqrt 2\quad ,
\end{equation}

\item {\em Periodic boundary conditions}

\qquad We first note that in this case the $S$-matrix constructed from the
transfer matrix $\widehat X_L"$ in Sect.~III is unitary (i.e. (\ref{37})
and (37a) are verified) only in the case where (\ref{38}) is obeyed
i.e. when chains 1 and 2 have identical random site-energies while the
random site energies on chain 3 are arbitrary.  This may be readily seen
by
expressing the averages of the transmission- and reflection coefficients
in (A.5) and (A.6) to second order in the site energies in terms of the
averages of (A.7-A.9) assuming that site energies
belonging to sites
$m\neq n$ on the same chain or on different ones are uncorrelated i.e.
$\langle \varepsilon_{im}\varepsilon_{jn}\rangle=0$.  In this way we find

\begin{eqnarray}\label{77}
\sum^3_{i,j=1}\left(\langle\mid t_{ij}^{--}\mid^2\rangle+\langle\mid
r_{ij}^{-+}\mid^2\rangle\right)
&=&
3+2N_L\left[(\langle c^2_n \rangle+\langle f^2_n \rangle-2\langle c_n f_n
\rangle) \right.\nonumber\\
&\quad& +(\langle d^2_n \rangle+\langle q^2_n \rangle-2\langle d_n q_n
\rangle)\nonumber\\
&\quad& +\left.(\langle g^2_n \rangle+\langle p^2_n \rangle-2\langle g_n
p_n \rangle)\right]\quad.
\end{eqnarray}

\noindent
Now, for this relation to be compatible with current conservation we
require

\begin{equation}\label{78}
c_n=f_n\;,\;d_n=q_n\;\text{and}\;g_n=p_n\quad ,
\end{equation}

\noindent
and from the definition (\ref{25}) it follows that the equalities
(\ref{78})
are fulfilled with (\ref{38}) i.e. for identical site energies in the
chains 1 and 2.  Only under this condition does the $S$-matrix for the
periodic 3-channel system represent a true scattering matrix

\qquad The explicit second order expressions for the averaged transmission-
and
reflection coefficients obtained from (A.5-A.9) and (\ref{25}), for random
gaussian site energies (\ref{27}) in the presence of the correlation
(\ref{38}) are

\begin{equation}\label{79}
\langle\mid t_{11}^{--}\mid^2\rangle=
1-{N_L\varepsilon^2_0\over 36}\left({5\over\sin^2k_1}+{8\over\sin k_1\sin
k_2}\right)\quad ,
\end{equation}

\begin{equation}\label{80}
\langle\mid t_{22}^{--}\mid^2\rangle=\langle\mid t_{33}^{--}\mid^2\rangle=
1-{N_L\varepsilon^2_0\over 36}\left({5\over\sin^2k_2}+{4\over\sin^2
k_2}+{4\over\sin k_1\sin k_2}\right)\quad ,
\end{equation}

\begin{equation}\label{81}
\langle\mid t_{12}^{--}\mid^2\rangle=\langle\mid t_{21}^{--}\mid^2\rangle=
\langle\mid t_{13}^{--}\mid^2\rangle=\langle\mid t_{31}^{--}\mid^2\rangle=
{2N_L\varepsilon^2_0\over 36}{1\over\sin k_1\sin k_2}\quad ,
\end{equation}

\begin{equation}\label{82}
\langle\mid t_{23}^{--}\mid^2\rangle=\langle\mid t_{32}^{--}\mid^2\rangle=
{2N_L\varepsilon^2_0\over 36}{1\over\sin^2 k_2}\quad ,
\end{equation}

\begin{equation}\label{83}
\langle\mid r_{11}^{-+}\mid^2\rangle=
{5N_L\varepsilon^2_0\over 36}{1\over\sin^2 k_1}\quad ,
\end{equation}

\begin{equation}\label{84}
\langle\mid r_{22}^{-+}\mid^2\rangle=\langle\mid r_{33}^{-+}\mid^2\rangle=
{5N_L\varepsilon^2_0\over 36}{1\over\sin^2 k_2}\quad ,
\end{equation}

\begin{eqnarray}\label{85}
\langle\mid r_{12}^{-+}\mid^2\rangle
&=&
\langle\mid r_{21}^{-+}\mid^2\rangle=
\langle\mid r_{13}^{-+}\mid^2\rangle=\langle\mid r_{31}^{-+}\mid^2\rangle=
{N_L\varepsilon^2_0\over 36}{2\over\sin k_1\sin k_2}\nonumber \\
\langle\mid r_{23}^{-+}\mid^2\rangle
&=&
\langle\mid r_{32}^{-+}\mid^2\rangle=
{2N_L\varepsilon^2_0\over 36}{1\over\sin^2 k_2}\quad ,
\end{eqnarray}

\noindent
The inverse localization length associated with the conductance (\ref{1})
obtained from (\ref{79}-\ref{82}) is

\begin{equation}\label{86}
{1\over L_c}={\varepsilon^2_0\over 216}
\left({5\over \sin^2 k_1}+{14\over \sin^2 k_2}+{8\over\sin k_1\sin k_2}\right)
\quad ,
\end{equation}

\noindent
where $k_1$ and $k_2$ are defined by (18b).  This expression which
proves localization in periodic 3-channel systems, reduces again to the 1D
result (\ref{52}) for $h=0$.  For small $h$ it is (with $E=2\cos k$)

\begin{equation}\label{87}
{1\over L_c}\simeq{\varepsilon^2_0\over 8\sin^2 k}
\left[1+{h^2\over 2\sin^2 k}\left(1+{70\over27}\cot^2k\right)\right]
\quad ,
\end{equation}

\noindent
which shows enhanced localization for weak interchain hopping.  On the
other hand for $\mid h\mid >> \mid E\mid$ (\ref{86}) becomes

\begin{equation}\label{88}
{1\over L_c}\simeq{5\varepsilon^2_0\over 216}{1\over 1-h^2}
\quad ,
\end{equation}

\noindent
which leads to increased localization lengths, $L_c>\xi_0$, for
$h^2<22/27$.  Thus the domain where the localization length in the
periodic 3-channel system is larger than the 1D-value is defined by

\begin{equation}\label{89}
\mid E\mid <<\mid h\mid <0.906\ldots\quad .
\end{equation}

\qquad Our exact microscopic results for two and three-channel systems
indicate
that in all case while in the three channel cases a weak interchain hopping
decreases the localization
length from its 1D-value.  For strong interchain hopping a similar
decrease
persists in the two-channel case while in the three-channel cases the
localization length increases with
respect to the 1D result over restricted domains of hopping parameters
defined by (\ref{76}) and (\ref{89}).  This suggests the possible
existence of quasi-metallic domains for the three-channel systems in some
ranges of length scales lying between their localization lengths and the
1D localization length (of the order of the mean free path $\ell$
\cite{10}) , in the
considered ranges of large hopping.  In contrast, we recall that for
many-channel random wires ($N>>1$) the metallic domain
exists over a wide range of mesoscopic lengths defined by (\ref{5}).

\end{enumerate}

\section{Appendix}

For briefness' sake we only discuss explicit expressions for transmission
amplitudes $t^{--}_{ij}$ and reflection amplitudes $r^{-+}_{ij}$ relating
to outgoing waves to the left of the disordered sample in the scattering
equations (\ref{36}) (see (\ref{33}) and (\ref{42})).

\begin{enumerate}
\item {\em Two-channel wires}

Identification of (\ref{33}) and (\ref{46}) yields

\begin{eqnarray*}
\mid t_{11}^{--}\mid^2
&=&
{\mid X_{44}\mid^2\over\mid\delta\mid^2}\;,\;
\mid t_{12}^{--}\mid^2={\mid X_{24}\mid^2\over\mid\delta\mid^2}\quad
,\nonumber \\
\mid t_{21}^{--}\mid^2
&=&
{\mid X_{42}\mid^2\over\mid\delta\mid^2}\;\text{and}\;
\mid t_{22}^{--}\mid^2={\mid X_{22}\mid^2\over\mid\delta\mid^2}\quad ,
\hspace{6cm}\text{(A.1)}
\end{eqnarray*}

\noindent
and, to lowest order in the random site energies, using (\ref{47}) and
(\ref{30}),

\begin{eqnarray*}
\mid r_{11}^{-+}\mid^2
&=&
\mid X_{21}\mid^2\;,\;
\mid r_{12}^{-+}\mid^2=
\mid X_{23}\mid^2\quad ,\nonumber\\
\mid r_{21}^{-+}\mid^2
&=&
\mid X_{41}\mid^2\;,\;
\mid r_{22}^{-+}\mid^2=
\mid X_{43}\mid^2\quad .
\hspace{5.8cm}\text{(A.2)}
\end{eqnarray*}

\noindent
>From the definition of $\delta$ in (\ref{47}) and the explicit form of the
elements $X_{ij}$ in (\ref{30}-\ref{31}), we obtain

\begin{eqnarray*}
\mid \delta\mid^2
&=&
1+\sum^{N_L}_{m,n=1}
[a_{1m}a_{1n}+a_{2m}a_{2n}+2b_{m}b_{n}\cos(m-n)(k_1-k_2)]\quad ,\nonumber \\
\mid X_{22}\mid^2
&=&
1+\sum_{m,n}a_{1m}a_{1n}\;,\;
\mid X_{44}\mid^2=
1+\sum_{m,n}a_{2m}a_{2n}\quad , \nonumber \\
\mid X_{12}\mid^2
&=&
\sum_{m,n} a_{1m}a_{2n}\cos 2(m-n)k_1\;,\;
\mid X_{24}\mid^2=
\sum_{m,n}b_{m}b_{n}\cos (m-n)(k_1-k_2)\quad ,
\quad\text{(A.3)}
\end{eqnarray*}

\begin{eqnarray*}
\mid r_{11}^{-+}\mid^2
&=&
\sum_{m,n} a_{1m}a_{1n}\cos 2(m-n)k_1\quad ,\nonumber \\
\mid r_{12}^{-+}\mid^2
&=&
\sum_{m,n} b_{m}b_{n}\cos (m-n)(k_1+k_2)\quad ,\nonumber \\
\mid r_{21}^{-+}\mid^2
&=&
\sum_{m,n} b_{m}b_{n}\cos (m-n)(k_1+k_2)\quad ,\nonumber \\
\mid r_{22}^{-+}\mid^2
&=&
\sum_{m,n} a_{2m}a_{2n}\cos 2(m-n)k_2\quad .
\hspace{6cm}\text{(A.4)}
\end{eqnarray*}

\item {\em Three channel wires}

The identification of (48.c) with the transmission matrix $\hat t^{--}$ in
(\ref{34}) yields

\begin{eqnarray*}
\mid t_{11}^{--}\mid^2
&=&
{\mid\beta_1\mid^2\over\mid\Delta\mid^2}\;,\;
\mid t_{12}^{--}\mid^2=
{\mid\beta_4\mid^2\over\mid\Delta\mid^2}\;,\;
\mid t_{13}^{--}\mid^2=
{\mid\beta_7\mid^2\over\mid\Delta\mid^2}\quad ,\nonumber \\
\mid t_{21}^{--}\mid^2
&=&
{\mid\beta_2\mid^2\over\mid\Delta\mid^2}\;,\;
\mid t_{22}^{--}\mid^2=
{\mid\beta_5\mid^2\over\mid\Delta\mid^2}\;,\;
\mid t_{23}^{--}\mid^2=
{\mid\beta_8\mid^2\over\mid\Delta\mid^2}\quad ,\nonumber \\
\mid t_{31}^{--}\mid^2
&=&
{\mid\beta_3\mid^2\over\mid\Delta\mid^2}\;,\;
\mid t_{32}^{--}\mid^2=
{\mid\beta_6\mid^2\over\mid\Delta\mid^2}\;,\;
\mid t_{33}^{--}\mid^2=
{\mid\beta_9\mid^2\over\mid\Delta\mid^2}\quad .
\hspace{6cm}\text{(A.5)}
\end{eqnarray*}

\noindent
Similarly, the reflection amplitudes $r^{-+}_{ij}$ are obtained by
identifying $\widehat S_1$, in (48.a) with (\ref{35}).  For analyzing
the reflection coefficients for weak disorder it suffices to find the
amplitudes to linear order in the site energies.  Such linear
contributions are associated with $\beta_j$-terms having a zeroth order
contribution (of modulus one).  From (48.a) and (48.e) we thus
obtain

\begin{eqnarray*}
\mid r_{11}^{-+}\mid^2
&=&
\mid Y_{21}\mid^2\;,\;
\mid r_{12}^{-+}\mid^2=
\mid Y_{23}\mid^2\;,\;
\mid r_{13}^{-+}\mid^2=
\mid Y_{25}\mid^2\quad ,\nonumber \\
\mid r_{21}^{-+}\mid^2
&=&
\mid Y_{41}\mid^2\;,\;
\mid r_{22}^{-+}\mid^2=
\mid Y_{43}\mid^2\;,\;
\mid r_{23}^{-+}\mid^2=
\mid Y_{45}\mid^2\quad ,\nonumber \\
\mid r_{31}^{-+}\mid^2
&=&
\mid Y_{61}\mid^2\;,\;
\mid r_{32}^{-+}\mid^2=
\mid Y_{63}\mid^2\;,\;
\mid r_{33}^{-+}\mid^2=
\mid Y_{65}\mid^2\quad .
\hspace{5cm}\text{(A.6)}
\end{eqnarray*}

\noindent
The explicit expressions of the $\mid\beta_j\mid^2$ and $\mid\Delta\mid^2$
in (A.5) (defined in (48.e,f) and of the reflection coefficients
(A.6) in terms of the transfer matrix elements in (\ref{32}) are given by

\begin{eqnarray*}
\mid\beta_1\mid^2
&=&
1+\sum_{m,n}
[a_{3m}a_{3n}+b_{2m}b_{2n}+2d_mq_n\cos(m-n)(k_3-k_2)]\quad ,\nonumber \\
\mid\beta_5\mid^2
&=&
1+\sum_{m,n}
[a_{1m}a_{1n}+a_{3m}a_{3n}+2g_mp_n\cos(m-n)(k_3-k_1)]\quad ,\nonumber \\
\mid\beta_9\mid^2
&=&
1+\sum_{m,n}
[a_{1m}a_{1n}+b_{2m}b_{2n}+2c_mf_n\cos(m-n)(k_2-k_1)]\quad ,\nonumber \\
\mid\beta_2\mid^2
&=&
\sum_{m,n}
f_{m}f_{n}\cos(m-n)(k_1-k_2)\;,\;
\mid\beta_4\mid^2=
\sum_m
c_mc_n\cos(m-n)(k_1-k_2)\quad ,\nonumber \\
\mid\beta_3\mid^2
&=&
\sum_{m,n}
p_{m}p_{n}\cos(m-n)(k_1-k_3)\;,\;
\mid\beta_7\mid^2=
\sum_{m,n}g_m g_n\cos(m-n)(k_1-k_3)\quad ,\nonumber \\
\mid\beta_6\mid^2
&=&
\sum_{m,n}
q_{m}q_{n}\cos(m-n)(k_2-k_3)\;,\;
\mid\beta_8\mid^2=
\sum_{m,n}d_m d_n\cos(m-n)(k_2-k_3)\quad ,
\hspace{1cm}\text{(A.7)}
\end{eqnarray*}

\begin{eqnarray*}
\mid\Delta\mid^2
&=&
1+\sum_{m,n}\left[a_{1m}a_{1n}+a_{3m}a_{3n}+b_{2m}b_{2n}+2g_mp_n\cos(m-n)(k_3-k_
4)\right.\nonumber \\
&+&
\left.2d_m q_n\cos(m-n)(k_3-k_2)+2c_m f_n\cos(m-n)(k_2-k_1)\right]
\quad ,
\hspace{3cm}\text{(A.8)}
\end{eqnarray*}

\begin{eqnarray*}
\mid r_{11}^{-+}\mid^2
&=&
\sum_{m,n}a_{1m}a_{1n}\;,\;
\mid r_{22}^{-+}\mid^2=
\sum_{m,n}b_{2m}b_{2n}\cos 2(m-n)k_2\quad ,\nonumber \\
\mid r_{33}^{-+}\mid^2
&=&
\sum_{m,n}a_{3m}a_{3n}\cos 2(m-n)k_3\;,\;
\mid r_{12}^{-+}\mid^2=
\sum_{m,n}c_{m}c_{n}\cos (m-n)(k_1+k_2)\quad ,\nonumber \\
\mid r_{13}^{-+}\mid^2
&=&
\sum_{m,n}g_{m}g_{n}\cos(m-n)(k_1+k_3)\;,\;
\mid r_{21}^{-+}\mid^2=
\sum_{m,n}f_{m}f_{n}\cos (m-n)(k_1+k_2)\quad ,\nonumber \\
\mid r_{23}^{-+}\mid^2
&=&
\sum_{m,n}d_{m}d_{n}\cos(m-n)(k_2+k_3)\;,\;
\mid r_{31}^{-+}\mid^2=
\sum_{m,n}p_{m}p_{n}\cos (m-n)(k_1+k_3)\quad ,\nonumber \\
\mid r_{32}^{-+}\mid^2
&=&
\sum_{m,n}q_{m}q_{n}\cos(m-n)(k_2+k_3)\quad ,
\hspace{7cm}\text{(A.9)}
\end{eqnarray*}

\noindent
where the double summations run from $m=1$ to $m=N_L$ and $n=1$ to
$n=N_L$.
The site-dependent tight-binding parameters $a_{1j}, a_{3j}, b_{2j},
c_j,d_j,f_j, g_j,p_j,q_j$ in the above expressions are defined by
(\ref{24})
for free boundary conditions and by (\ref{25}) for periodic boundary
conditions.  Likewise the wavenumbers $k_1,k_2,k_3$ are given by
(18.a)
and by (18.b) for free- and for periodic boundary conditions,
respectively.
\end{enumerate}


\newpage
\begin{references}
\bibliographystyle{unsrt}
\bibitem{1} Y.~Imry and R.~Landauer, Rev. Mod. Phys. {\bf71}, 306 (1999).
\bibitem{2} S.~Datta, Electronic transport in Mesoscopic Systems
(Cambridge University Press, 1995).
\bibitem{3} B.J.~van Wees {\it et al.}, Phys. Rev. Letters {\bf60}, 848
(1988); D.A.~Wharam {\it et al.}, J. Phys. C: Solid
State Phys. {\bf 21}, 209 (1988); see also H.~Van Houten and
C.W.J.~Beenakker, Phys. Today {\bf 49}, No~7, 22 (1996).
\bibitem{4} For recent reviews see A.D. Stone, P.A. Mello, K.A. Muttalib
and J.L.~Pichard in Mesoscopic
Phenomena in Solids ed. by B.L.~Altshuler, P.A.~Lee and R.A.~Webb (North
Holland, Amsterdam, 1991); C.W.J.~Beenakker, Rev.
Mod. Phys. {\bf 69}, 760 (1997).
\bibitem{5} D.J.~Thouless, Phys. Rev. Letters {\bf39}, 1167 (1977).
\bibitem{6} Y.~Imry, Introduction to Mesoscopic Physics (Oxford
University, London 1995).
\bibitem{7} Y. Imry, Europhys. Lett. {\bf1}, 249 (1986).
\bibitem{8} O.N.~Dorokhov, Zh. Eksp. Teor. Fiz. {\bf85}, 1040 (1983) [Sov.
Phys. JETP {\bf 58}, 606].
\bibitem{9} P.W.~Anderson, D.J.~Thouless, E.~Abrahams, and D.S.~Fisher,
Phys. Rev. B {\bf22}, 3519 (1980).
\bibitem{10} D.J.~Thouless, J. Phys. C: Solid State Phys. {\bf6}, 249
(1973).
\bibitem{11} See e.g. M.~Janssen, Phys. Rep. {\bf295}, 1 (1998).
\bibitem{12} O.N.~Dorokhov, Solid State Com. {\bf44}, 915 (1982).
\bibitem{13} D.J.~Thouless, in Ill Condensed Matter, edited by R.~Balian,
R.~Maynard, and G.~Toulouse (North Holland,
Amsterdam, 1979).
\bibitem{14} See e.g. D Mailly, C. Chapelier, and A. Benoit, Phys. Rev.
Lett. {\bf 70}, 2020 (1993).
\bibitem{15} R.~Johnston and H.~Kunz, J. Phys. C: Solid State Phys.
{\bf16}, 3895 (1983).
\bibitem{16} V.I.~Oseledec, Trans. Moscow Math. Soc. {\bf19}, 197 (1968).
\bibitem{17} V.N.~Tutubalin, Theor. Prob. Appl. {\bf13}, 65 (1968);
A.D.~Vister, ibid. {\bf 15}, 667 (1970).
\bibitem{18} See also the reviews on localization in A.~Crisanti,
G.~Paladin, A.~Vulpiani, Products of Random Matrices
(Springer, Berlin, 1993); M.~Janssen, O.~Viehweger, V.~Fastenrath, and
J.~Hajdu, Introduction to the Theory of the Integer
Quantum Hall Effect (VCH, New York, 1994).
\bibitem{19} J.B.~Pendry, Adv. Phys. {\bf43}, 461 (1994).
\bibitem{20} P.~Erd\"{o}s and R.C.~Herndon, Adv. Phys. {\bf31}, 65 (1982).
\bibitem{21} P.W.~Anderson and P.A.~Lee, Supp. Prog. Theor. Phys. {\bf69},
212 (1980).

\end{references}
\end{document}